\documentclass[12pt]{article}
\usepackage{epsf}
\usepackage[dvips]{graphicx}

\renewcommand{\thefootnote}{\#\arabic{footnote}}
\begin{document}

\newcommand{\bear}{\begin{array}}  \newcommand{\eear}{\end{array}}
\newcommand{\bea}{\begin{eqnarray}}  \newcommand{\eea}{\end{eqnarray}}
\newcommand{\beq}{\begin{equation}}  \newcommand{\eeq}{\end{equation}}
\newcommand{\bef}{\begin{figure}}  \newcommand{\eef}{\end{figure}}
\newcommand{\bec}{\begin{center}}  \newcommand{\eec}{\end{center}}
\newcommand{\non}{\nonumber}  \newcommand{\eqn}[1]{\beq {#1}\eeq}
\newcommand{\lmk}{\left(}  \newcommand{\rmk}{\right)}
\newcommand{\lkk}{\left[}  \newcommand{\rkk}{\right]}
\newcommand{\lhk}{\left \{ }  \newcommand{\rhk}{\right \} }
\newcommand{\del}{\partial}  \newcommand{\abs}[1]{\vert{#1}\vert}
\newcommand{\vect}[1]{\mbox{\boldmath${#1}$}}
\newcommand{\bib}{\bibitem} \newcommand{\new}{\newblock}
\newcommand{\la}{\left\langle} \newcommand{\ra}{\right\rangle}
\newcommand{\bfx}{{\bf x}} \newcommand{\bfk}{{\bf k}}
\newcommand{\gtilde} {~ \raisebox{-1ex}{$\stackrel{\textstyle >}{\sim}$} ~} 
\newcommand{\ltilde} {~ \raisebox{-1ex}{$\stackrel{\textstyle <}{\sim}$} ~}
\newcommand{\gtrsim}{ \mathop{}_{\textstyle \sim}^{\textstyle >} }
\newcommand{\lesssim}{ \mathop{}_{\textstyle \sim}^{\textstyle <} }

\renewcommand{\thefootnote}{\fnsymbol{footnote}}
\setcounter{footnote}{0}
\begin{titlepage}

\def\thefootnote{\fnsymbol{footnote}}

\begin{center}

\hfill HIP-2003-63/TH\\
\hfill TU-705\\
\hfill RESCEU-53/03\\
\hfill hep-ph/0312094\\
\hfill December, 2003\\

\vskip .5in

{\Large \bf

Can MSSM Particle be the Inflaton?

}

\vskip .45in

{\large
Shinta Kasuya$^{(a)(b)}$,
Takeo Moroi$^{(c)}$ and
Fuminobu Takahashi$^{(d)}$
}

\vskip .3in

{\em $^{(a)}$
Helsinki Institute of Physics\\
P.O. Box 64, FIN-00014 University of Helsinki, Finland
}

\vskip .1in

{\em $^{(b)}$
Department of Information Science, Kanagawa University\\
Kanagawa 259-1293, Japan
}

\vskip .2in

{\em 
$^{(c)}$Department of Physics,  Tohoku University,
Sendai 980-8578, Japan
}

\vskip .2in

{\em
$^{(d)}$Research Center for the Early Universe, School of Science,
University of Tokyo\\
Tokyo 113-0033, Japan
}

\end{center}

\vskip .4in

\begin{abstract}

We consider the possibility of using one of the $D$-flat directions in
the minimal supersymmetric standard model (MSSM) as the inflaton.  We
show that the flat direction consisting of (first generation) left-
and right-handed up-squarks as well as the up-type Higgs boson may
play the role of the inflaton if dominant part of the up-quark mass is
radiatively generated from supersymmetric loop diagrams.  We also
point out that, if the $R$-parity violating Yukawa coupling is of
$O(10^{-7})$, $R$-odd $D$-flat directions may be another possible
candidate of the inflaton.  Such inflation models using $D$-flat
directions in the MSSM are not only testable with collider experiments
but also advantageous to resolve the problem how the inflaton reheats
the universe.

\end{abstract}

\end{titlepage}

\renewcommand{\thepage}{\arabic{page}}
\setcounter{page}{1}
\renewcommand{\thefootnote}{\#\arabic{footnote}}
\setcounter{footnote}{0}

Inflation \cite{inflation} is now one of the most important ideas in
cosmology.  Inflation not only solves the serious horizon and flatness
problems but also provides a viable scenario of generating the origin
of cosmic density fluctuations.\footnote
{Another possibility of generating cosmic density fluctuations may be
to adopt the ``curvaton'' scenario \cite{curvaton}.  Here, we do not
consider such a possibility and assume that the cosmic density
fluctuations are totally generated from the primordial fluctuation of
the inflaton.}
In particular, precise measurement of the anisotropy of the cosmic
microwave background (CMB) by the Wilkinson Microwave Anisotropy Probe
(WMAP) suggests that the primordial density fluctuations are almost
purely adiabatic and scale-invariant \cite{WMAP}, which are
predictions of (some classes of) inflationary models. In the
inflationary models, a scalar field, called ``inflaton,'' is
introduced to realize the inflationary epoch of the universe.  During
inflation, potential energy of the inflaton gives 
the dominant part of the energy density.  In order to realize the
quasi de Sitter universe with the potential energy of the inflaton,
its kinetic energy should be much smaller than the potential energy
during inflation.  Consequently, we are led to the paradigm of
slow-roll inflation, where inflation is driven by the potential energy
of slowly evolving scalar field.

It is non-trivial to find a scalar field which satisfies the slow-roll
condition, and from the particle-physics point of view, it is
important to find good candidates of the inflaton.  In particular, it
is interesting to ask if the inflaton can be observed at high-energy
collider experiments in any scenario of inflation.  In many classes of
models, however, a new scalar field is introduced as the inflaton.
Usually, such scalar fields do not belong to the standard model and
have very weak interactions with the standard-model particles, which
makes it very difficult to find and study the inflaton with collider
experiments. Moreover, since the interactions between such inflaton
and standard-model particles cannot be determined, it must be given
{\it ad hoc} by hand, which obscures the thermal history of the
universe. Thus the reheating temperature, for instance, cannot surely
be estimated. In order to construct a testable and economical model of
inflation, it is desirable to find a candidate of the inflaton in the
scalar fields which are in some sense familiar to us.  In fact, in the
standard model, the only scalar field is the Higgs boson.  As we will
see later, however, it is known that the Higgs boson in the standard
model cannot be the inflaton since its (quartic) coupling is too large
to generate cosmic density fluctuations that are consistent with the
observations. In addition, in the standard model, the potential of the
Higgs boson is significantly affected by radiative corrections and
hence we cannot expect flat enough potential which is crucial to cause
inflation long enough. Thus, we should conclude that it is impossible
to find a viable candidate of the inflaton in the standard-model
fields. In fact, the second point, effects of radiative corrections to
the inflaton potential, is in general a serious problem in
constructing inflation models.

In order to control the radiative corrections, it is often the case
that inflation models are considered in supersymmetric framework;
indeed, in supersymmetric models, quadratic divergences cancel out
between bosonic and fermionic loops and the flatness of the potential
can be guaranteed.  Thus, in considering the inflation, supersymmetry
is likely to play very important roles, and in this letter, we adopt
(low-energy) supersymmetry.  If we supersymmetrize the standard model,
various scalar particles are introduced as superpartners of
quarks and leptons.  Those scalars may play the role of inflaton. It
is worth noting that, if this is the case, the reheating processes
into the standard-model particles are obvious, which makes it possible
to study both the inflationary and thermal history of the universe
only with the low-energy ``known'' physics.  Actually
there was an early attempt to build inflation models along this 
line~\cite{Connors:1988yx}. However, as discussed later, it contained some 
difficulty in producing the right amount of density fluctuations, so the authors of 
Ref.~\cite{Connors:1988yx} had to introduce additional mini-inflation.
Here we would like to pursue another possibility.

In this letter, we consider the possibility of using scalar fields
in the minimal supersymmetric standard model (MSSM) as the inflaton
(which is denoted as $\phi$ hereafter).  In particular, we will
discuss that the $D$-flat direction consisting of first-generation
left- and right-handed up-type squarks as well as the up-type Higgs
boson may play the role of the inflaton if the up-quark mass is
radiatively generated.  We will also see that, if the $R$-parity
violating Yukawa coupling is of $O(10^{-6}-10^{-7})$, $R$-parity
violating $D$-flat directions may also be the inflaton.

We start with discussing possible scenario of inflation within the
MSSM.  In the framework of the slow-roll inflation, the amplitude of
the inflaton field varies during and after the inflation.  The change
of the inflaton amplitude is typically of $O(M_*)$, where $M_*\simeq
2.4\times 10^{18}\ {\rm GeV}$ is the reduced Planck scale.\footnote
{We consider neither the so-called small-field models (for example,
new inflation) nor hybrid models within the MSSM. Generally speaking,
both require the vacuum to be such that the standard-model gauge
symmetries are spontaneously broken, which is hardly justified.  Here
and hereafter we concentrate on the so-called large-field models ({\it
i.e.}, chaotic inflation). }
Thus, inflaton originating from the MSSM sector should have an
amplitude of $O(M_*)$ during inflation.  Consequently, we are
forced to consider the chaotic inflation \cite{PLB129-177} within the
MSSM assuming that the MSSM field corresponding to the inflaton has an
amplitude of the order of the Planck scale during inflation.
Later we will discuss how to realize such a large field value.

Now, we discuss the observational constraints on the inflaton
potential.  In the simplest scenario, the cosmic density fluctuations
are parameterized by the curvature perturbation ${\cal R}$ which
depends on the inflaton potential $V$ as
\begin{eqnarray}
{\cal R} (k) = 
\left[ 
\frac{H_{\rm inf}}{2\pi}
\frac{3H_{\rm inf}^2}{V'}
\right]_{k=a H_{\rm inf}},
\end{eqnarray}
where $H_{\rm inf}$ is the expansion rate $H$ of the universe during
inflation, which is related to the potential energy of the inflaton
during inflation as $H_{\rm inf}^2 = V/3M_*^2$, while $V'$ is the
derivative of the inflaton potential with respect to the inflaton
field.  Here, notice that this quantity is evaluated at the time when
the fluctuation exits the horizon during inflation; $k$ and $a$ denote
the wave-number (for the comoving coordinate) and scale factor,
respectively.

   From the measurements of the CMB anisotropies (as well as from other
observations), we can obtain constraints on ${\cal R} (k)$.  The most
important constraint is on its normalization.  Due to the fact that
$\Delta T/T\sim O(10^{-5})$, the typical size of ${\cal R}$ is also
constrained to be $O(10^{-5})$.  More precisely, the WMAP
team~\cite{WMAP} estimated it as $|{\cal R}(k)|^2 = 2.95 \times
10^{-9} A$ with $A=0.9 \pm 0.1$ at $k=0.05 {\rm ~Mpc}^{-1}$, assuming
power-law $\Lambda$CDM model. If the inflaton potential has the
parabolic form $V=\frac{1}{2}M_\phi^2 \phi^2$, the inflaton mass is
required to be $M_\phi\sim O(10^{13}\ {\rm GeV})$.  Obviously, such a
heavy scalar field does not exist in the MSSM.

We consider the next possibility where the inflaton potential is
quartic:
\begin{eqnarray}
V = \frac{1}{4} \lambda \phi^4,
\label{phi^4}
\end{eqnarray}
where $\lambda$ is a dimensionless coupling constant.  Then, the
normalization of the primordial density fluctuation requires
$\lambda\sim O(10^{-13})$ as follows. The value of the inflaton field
is related to the $e$-folding number $N$ as $\phi\simeq\sqrt{8 N}
M_*$.  (See Eq.~(\ref{eq:efold}) below). Thus the curvature
perturbation at the horizon exit is evaluated as ${\cal R}\simeq 0.3
\sqrt{\lambda} N^{3/2}$. Equating this with the WMAP result, we obtain
$\lambda\sim 10^{-13}$, where $N \sim 60$ is used.  Before studying
the MSSM cases, we should comment on the non-supersymmetric case; from
this constraint, it is obvious that the Higgs boson cannot play the
role of the inflaton since, if the quartic coupling of the Higgs boson
is as small as $O(10^{-13})$, Higgs mass becomes $O(10^{-4}\ {\rm
GeV})$ which is unacceptably smaller than the present experimental
bound.

In the supersymmetric case, such a small coupling for quartic
interaction cannot be realized if the potential is lifted by the gauge
$D$-term interactions, since, if so, the coupling constant $\lambda$
becomes of the order $O(g^2)$ where $g$ is the gauge coupling constant
in the standard model.  Therefore we focus our attention on the
$D$-flat directions. For $D$-flat directions, we have to be more
careful since behaviors of the potential depend on which flat
direction we consider.  In the MSSM, Yukawa interactions exist in the
superpotential to generate the fermion masses.  Such Yukawa
interactions lift some of the $D$-flat directions.  In addition, we
can also find several $D$-flat directions which are not affected by
the Yukawa interactions associated with the fermion masses; without
$R$-parity violation, such $D$-flat directions are only lifted by the
effects of supersymmetry breaking.\footnote
{Here, we assume that coefficients of non-renormalizable terms are
suppressed enough to be neglected.  This may be explained by the
$R$-symmetry, assigning $R$-charge $\frac{2}{3}$ to each MSSM chiral
superfields.}
(See Ref.~\cite{Gherghetta} for the details.)

We first consider the $D$-flat direction lifted by the $R$-parity
conserving Yukawa interactions.  $D$-flat directions are parameterized
by gauge invariant monomial of the superfields.  We denote the MSSM
superfields as $Q({\bf 3}, {\bf 2}, \frac{1}{6})$, $U({\bf 3^*}, {\bf
1}, -\frac{2}{3})$, $D({\bf 3^*}, {\bf 1}, \frac{1}{3})$, $L({\bf 1},
{\bf 2}, -\frac{1}{2})$, $E({\bf 1}, {\bf 1}, 1)$, $H_u({\bf 1}, {\bf
2}, \frac{1}{2})$, and $H_d({\bf 1}, {\bf 2}, -\frac{1}{2})$, where we
show the quantum numbers for the $SU(3)_{\rm C}\times SU(2)_{\rm
L}\times U(1)_{\rm Y}$ gauge group in the parentheses.  Then, the
relevant part of the MSSM superpotential is given by
\begin{eqnarray}
W = 
[Y_U]_{ij} Q_i U_j H_u
+ [Y_D]_{ij} Q_i D_j H_d
+ [Y_E]_{ij} L_i E_j H_d,
\label{W}
\end{eqnarray}
where $i$ and $j$ are generation indices.  If we consider the $D$-flat
directions represented as $Q_iU_jH_u$, $Q_iD_jH_d$, and $L_iE_jH_d$,
those $D$-flat directions acquire quartic potential due to the Yukawa
interactions, as given in Eq.\ (\ref{phi^4}).  In order to relate the
Yukawa coupling constants to $\lambda$ in Eq.~(\ref{phi^4}), we
express those $D$-flat directions with a complex scalar field $\Phi$,
for example, 
\beq
\label{eq:fdquh}
Q_i = \frac{1}{\sqrt{3}} \left(\bear{c}
\Phi \\
0
\eear\right),~~~
U_j = \frac{1}{\sqrt{3}} \Phi,~~~
H_u = \frac{1}{\sqrt{3}} \left(\bear{c}
0\\
\Phi
\eear\right).
\eeq
Self quartic coupling constants of those flat directions, denoted as
$\lambda_{Q_iU_jH_u}$, $\lambda_{Q_iD_jH_d}$, and
$\lambda_{L_iE_jH_d}$, respectively, are then given by
\begin{eqnarray}
\lambda_{Q_iU_jH_u} = \frac{1}{3} \left| [Y_U]_{ij} \right|^2,~~~
\lambda_{Q_iD_jH_d} = \frac{1}{3} \left| [Y_D]_{ij} \right|^2,~~~
\lambda_{L_iE_jH_d} = \frac{1}{3} \left| [Y_E]_{ij} \right|^2,
\end{eqnarray}
where we defined $\phi \equiv \sqrt{2}{\rm Re\,} \Phi$.  Thus, if one
of these coupling constants is of $O(10^{-13})$, we may have inflaton
candidates within the MSSM particles.

If the fermion masses totally originate from the superpotential given
in Eq.\ (\ref{W}), Yukawa coupling constants are evaluated as
\begin{eqnarray}
[\hat{Y}_U]_{ii} = \frac{m_{u_i}}{v\sin\beta},~~~
[\hat{Y}_D]_{ii} = \frac{m_{d_i}}{v\cos\beta},~~~
[\hat{Y}_E]_{ii} = \frac{m_{e_i}}{v\cos\beta},~~~
\label{YukawavsM}
\end{eqnarray}
where $\hat{Y}_U$, $\hat{Y}_D$, and $\hat{Y}_E$ are Yukawa matrices in
the diagonalized basis, while $m_{u_i}$, $m_{d_i}$, and $m_{e_i}$ are
fermion masses for corresponding Yukawa couplings.  In addition,
$v\equiv\sqrt{\langle H_u^2\rangle + \langle H_d^2\rangle}\simeq 174\ 
{\rm GeV}$ while $\tan\beta\equiv \langle H_u\rangle/\langle
H_d\rangle$.  Yukawa coupling constants for the second and third
generation quarks and leptons are much larger than $\sim 10^{-6}$, and
hence we consider the possibility of using the first generation
squarks and/or sleptons as the inflaton. (Hereafter, we consider only
the first generation squarks and sleptons, and drop the generation
indices for simplicity unless otherwise mentioned.)  If we estimate
the Yukawa coupling constants for the up and down-quarks as well as
electron using Eq.\ (\ref{YukawavsM}), we obtain
\begin{eqnarray}
y_u \simeq \frac{8.6 \times 10^{-6}}{\sin\beta} 
\times \left( \frac{m_u}{1.5\ {\rm MeV}} \right),~~~
y_d \simeq \frac{2.9\times 10^{-5}}{\cos\beta} 
\times \left( \frac{m_d}{5\ {\rm MeV}} \right),~~~
y_e \simeq \frac{2.9 \times 10^{-6}}{\cos\beta}.
\end{eqnarray}
In the MSSM, it is often the case that, in order to evade the
Higgs-mass constraint, relatively large value of $\tan\beta$ is
required \cite{MSSMHiggs}.  Then, $y_d$ and $y_e$ are likely to be
larger than $10^{-6}$.  In fact, if we adopt the up-quark mass
$m_u=1.5 - 4.5 \ {\rm MeV}$ \cite{PRD66-010001},\footnote
{Here, we do not consider renormalization group running which
suppresses the fermion masses at higher energy scale.  Even if the
renormalization group effects are taken into account, our discussions
are qualitatively unchanged.}
we obtain $y_u\ge 8.6 \times 10^{-6}$, leading to $\lambda_{QUH_u} \ge
2.5 \times 10^{-11}$, and hence even the up-squark cannot play the
role of the inflaton.\footnote{
The quark mass ratios are constrained by imposing a limit on
next-to-leading order corrections in the chiral perturbation theory.
If the corrections become sizable, the up quark mass can be much
smaller than the lower limit $\sim 1.5$
MeV~\cite{Kaplan:1986ru}. Then, $y_{u}$ can be as small as $\sim
O(10^{-7})$.  In this case, we do not have to consider the radiatively
induced up-quark mass as we will discuss in the following.
}
One might well give up the simple $\lambda \phi^4$ model and add another
mini-inflation to reconcile the predicted magnitude of density fluctuations with
the observed one~\cite{Connors:1988yx}. However, here 
we would like to stick to the $\lambda \phi^4$ model,
since it  predicts the density fluctuations with very small uncertainty as shown later.

\begin{figure}[t]
    \begin{center}
        \scalebox{0.75}{\includegraphics{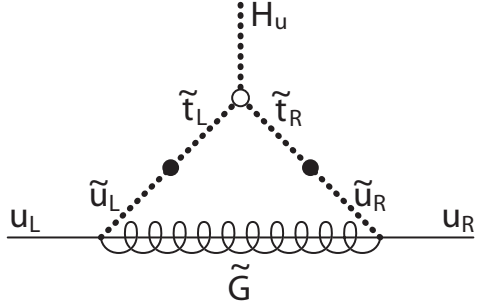}}
        \caption{Loop diagram contributing to the up-quark 
        mass in the MSSM.  Black dots represent the insertion of the
        off-diagonal elements of the squark mass matrices while the
        vertex with the open circle is from the insertion of the
        trilinear coupling $A_t$.}
        \label{fig:dyloop}
    \end{center}
\end{figure} 

So far, we have assumed that the Yukawa interactions in the
superpotential are the only sources of the fermion masses.  In the
MSSM, however, radiative corrections due to the supersymmetric loop
also affect the fermion masses.  If such an effect is large enough to
explain the dominant part of the fermion mass, Yukawa coupling
constant in the superpotential can be smaller than the naive
expectation.  Indeed, for the up-quark mass, contribution from the
squark-gluino loop diagram may become sizable with non-vanishing
off-diagonal elements of the left- and right-handed squark mass
matrices, as given in Fig.\ \ref{fig:dyloop}.  Relevant part of the
supersymmetry breaking terms contributing to this diagram is given by
\begin{eqnarray}
    {\cal L} &=& 
    \Delta m^2_{\tilde{u}_L\tilde{t}_L} 
    \tilde{u}_L \tilde{t}_L^*
    + \Delta m^2_{\tilde{u}_R\tilde{t}_R} 
    \tilde{u}_R \tilde{t}_R^*
    + A_t \tilde{t}_L \tilde{t}_R H_u
    + \frac{1}{2} m_{\tilde{G}} \tilde{G}\tilde{G} + {\rm h.c.},
\end{eqnarray}
where $\tilde{u}_{L,R}$ and $\tilde{t}_{L,R}$ are left- and
right-handed up- and top-squarks, respectively, while $\tilde{G}$
denotes the gluino.  Hereafter, we approximate that the masses of the
squarks are all degenerate for simplicity, and we define
\begin{eqnarray}
    x \equiv \frac{m_{\tilde{G}}^2}{m_{\tilde{q}}^2},~~~
    \delta^{(L,R)}_{13} \equiv
    \frac{\Delta m^2_{\tilde{u}_{L,R}\tilde{t}_{L,R}}}
    {m^2_{\tilde{q}}},~~~
    a_t \equiv \frac{A_t}{y_tm_{\tilde{q}}},
\end{eqnarray}
where $m_{\tilde{q}}$ is the squark mass, and $y_t$ is the top Yukawa
coupling constant.  Then, the loop contribution to the up-quark mass
is given by
\begin{eqnarray}
    m_u^{\rm (loop)} = \frac{1}{36\pi^2} g_3^2 y_t
    a_t \delta^{(L)}_{13} \delta^{(R)}_{13} \langle H_u \rangle
    \frac{x^{1/2} (x^3-6x^2+3x+2+3x\ln x)}{(x-1)^4},
\label{mu(loop)}
\end{eqnarray}
where $g_3$ is the $SU(3)_{\rm C}$ gauge coupling constant.

Size of the off-diagonal elements of the squark mass matrices are
constrained from flavor changing processes.  In particular, for
$\delta^{(L)}_{13}$, one finds $\delta^{(L)}_{13}<4.6\times 10^{-2}$
($9.8\times 10^{-2}$, $2.3\times 10^{-1}$) for $x=0.3$ ($1.0$ and
$4.0$) and $m_{\tilde{q}}=500\ {\rm GeV}$ \cite{NPB477-321}.\footnote
{Constraint on $\delta^{(L)}_{13}$ here is from the mass difference of
the neutral $B_d$ mesons.  Such a constraint is obtained from the
basis where the down-type squarks are diagonalized.  Two basis are
related by the Kobayashi-Maskawa matrix, but we neglect the difference
of the two basis because the change of the constraint is small enough
to be neglected.}
Constraint on $\delta^{(R)}_{13}$ is, on the other hand, not available
since we do not have precise measurement of the flavor changing decay
of the top quark.  Thus, the combination
$\delta^{(L)}_{13}\delta^{(R)}_{13}$ can be of $O(0.1)$ without
conflicting experimental constraints.\footnote
{If the supersymmetry breaking parameters have large complex phases,
too large neutron electric dipole moment may be induced in this model.
Here, we consider the case where the phases in the supersymmetry
breaking parameters are small enough.  We also assume that the mixing
parameters for the first and second generation squarks are suppressed
in order to evade the constraints from the $K^0$-$\bar{K}^0$ mixing.}
Then, for example for the case
where $m_{\tilde{G}}\simeq m_{\tilde{q}}$, we obtain
\begin{eqnarray}
    m_u^{\rm (loop)} \simeq
    \frac{1}{72\pi^2} g_3^2 y_t a_t
    \delta^{(L)}_{13} \delta^{(R)}_{13} \langle H_u \rangle
    \simeq 3.6\ {\rm MeV} \times y_t a_t \sin\beta
    \left( 
        \frac{\delta^{(L)}_{13} \delta^{(R)}_{13}}{10^{-2}}
    \right).
\end{eqnarray}
Thus, the up-quark mass of the size of a few MeV can be generated from
the loop effect with reasonable value of $a_t\sim O(1)$.  Then, the
up-Yukawa coupling constant in the superpotential can be of
$O(10^{-6}-10^{-7})$.  Notice that the loop-induced up-quark mass is
not suppressed even if the masses of the superparticles become much
larger than the electroweak scale (as far as all the supersymmetry
breaking parameters are of the same order).  Thus, if one wishes to
push up $m_u^{\rm (loop)}$ without affecting flavor-changing
processes, one possibility is to assume (relatively) large masses for
the superparticles.  We can also consider radiative correction to the
down quark mass.  In this case, $A_t$ should be replace by $y_b\mu_H$
in the calculation, where $y_b$ and $\mu_H$ are bottom-quark Yukawa
coupling and supersymmetric Higgs mass, respectively, and the
generation mixings are provided by the down-type squark mass matrices.
Thus, radiative correction to the down-quark mass may become sizable
when $\tan\beta$ is large.  Constraints on the off-diagonal elements
of the down-type scalar quark mass matrices are, however, more
stringent than those on the up-type scalar quark mass matrices.  Thus,
it is more difficult to radiatively generate the down-quark mass.

The above arguments open a window to use the $D$-flat direction
$QUH_u$ as the inflaton.  Thus, hereafter, we concentrate on the
evolution of the universe for the case where $D$-flat direction
$QUH_u$ plays the role of the inflaton, assuming that the dominant
part of the up-quark mass is from the radiative correction.

The potential of $QUH_u$ is now given by
\beq
V(\phi) = \frac{1}{2} m_{\phi}^2 \phi^2 + \frac{\lambda}{4} \phi^4,
\eeq
where $m_{\phi} \sim O(100\ {\rm GeV})$ is the soft supersymmetry
breaking mass. These two terms become comparable at $\phi_m \equiv
\sqrt{2/\lambda} \,m_{\phi}$.  Since the mass term can be neglected
for large value of $\phi \gg \phi_m$, its dynamics is just the same as
the well-known $\lambda \phi^4$ model.  First let us briefly review
the $\lambda \phi^4$ model and discuss the observational constraint on
that.  In this model, the slow roll parameters are given by
\begin{eqnarray}
\epsilon \equiv \frac{1}{2} M_*^2 \left( \frac{V'}{V} \right)^2
= 8\frac{M_*^2}{\phi^2},~~~
\eta \equiv M_*^2 \frac{V''}{V} = 12\frac{ M_*^2}{\phi^2},
\label{eq:sr_parameter}
\end{eqnarray}
where we substituted the quartic potential Eq.~(\ref{phi^4}). When the
inflaton $\phi$ becomes equal to $\phi_{\rm end} \equiv 2\sqrt{3}
M_*$, the slow-roll condition breaks down, leading to the end of
inflationary epoch.  The $e$-folding number $N$ is given by
\bea
\label{eq:efold}
N \equiv  \int_t^{t_{\rm end}} H dt 
\simeq \int_{\phi_{\rm end}}^\phi \frac{3 H^2}{V'} d\phi
= \frac{1}{8M_*^2}(\phi^2-\phi_{\rm end}^2),
\eea
where we used the slow-roll approximation in the second equality.

In order to impose observational constraints on inflation models, it
is necessary to evaluate several parameters that characterize the
density fluctuations generated during inflation. They are scalar
spectral index $n_s$, its running $d n_s/d\ln{k}$, and
tensor-to-scalar ratio $r$, apart from the normalization that
determines the value of $\lambda$. These are related to the
$e$-folding number as
\bea
n_s = 1-\frac{3}{N+\frac{3}{2}},~~~
\frac{d n_s}{d \ln{k}} = -\frac{3}{(N+\frac{3}{2})^2},~~~
r = \frac{16}{N+\frac{3}{2}}.
\label{eq:sr-N}
\eea
Thus we must precisely evaluate $N$ in order to make a comparison
between predictions and observational results.  Since the value of $N$
necessary to solve the horizon and flatness problems depends on the
thermal history of the universe, we must specify the reheating
processes. Fortunately, however, it is possible to determine the
$e$-folding number as $N \simeq 64$ with small uncertainty in the case
of $\lambda \phi^4$ model.  The reason is that the energy density of
inflaton oscillation decreases just like radiation after the end of
inflation, which applies only to the quartic model. The explicit
expression for $N$ is given by~\cite{Liddle:1993fq}
\beq
N = 62-\ln{\left(\frac{k}{a_0 H_0}\right)}
- \ln{\left(\frac{10^{16}\ {\rm GeV}}{V_k^{1/4}}\right)}
+ \ln{\left(\frac{V_k^{1/4}}{V_{\rm end}^{1/4}}\right)},
\label{eq:efold_exp}
\eeq
where $V$ represents the potential energy of the inflaton, and the
subscripts ``0,'' ``$k$,'' and ``end'' are for variables at present,
at the time when the fluctuation with the wave-number $k$ exits the
horizon, and at the end of the inflation, respectively.  The right
hand side of Eq.~(\ref{eq:efold_exp}) gives $N \simeq 64$ for $k=a_0
H_0$, irrespective of the details of the reheating. Here we have
assumed that the reheating processes complete before the amplitude of
the inflaton becomes smaller than $\phi_m$, which will be justified
below.  Substituting $N \simeq 64$ into Eqs.~(\ref{eq:sr-N}), we
obtain $n_s \simeq 0.95$ and $r \simeq 0.24$ with negligible running
of $n_s$.

Constraints on the inflationary models driven by a single slow-rolling
scalar field from the recent observations including the WMAP have been
studied extensively~\cite{Peiris:2003ff,Barger:2003ym,Kinney:2003uw}.
Although Ref.~\cite{Peiris:2003ff} claimed that $\lambda \phi^4$ model
lies in the region marginally excluded by the WMAP data in combination
with smaller scale CMB and large scale structure survey data, more
detailed analyses~\cite{Barger:2003ym,Kinney:2003uw} showed that the
model cannot be excluded by the WMAP data alone for $N \gtrsim 40$. In
addition, the recent systematic study using both the Sloan Digital Sky
Survey and WMAP demonstrated that $\lambda \phi^4$ model is still
allowed~\cite{Tegmark:2003ud}. Thus there is no reason to disregard
$\lambda \phi^4$ model at present, and one could well contend that the
model is of much interest since it is on the edge.

The next discussion concerns the decay processes of the
inflaton. After inflation ends, the inflaton oscillates around its
origin until the decay completes. In the usual chaotic inflation
models, the reheating process proceeds through nonperturbative
particle creation (preheating) \cite{preh}. During the oscillation,
the particles coupled to the inflaton are produced when the so-called
adiabatic condition is violated, {\it i.e.}, $\dot{\omega}/\omega^2
\gtrsim 1$, where $\omega$ is the effective frequency of the produced
particles. In our model, the inflaton field is complex, and its
nontrivial trajectory in the potential may lessen the efficiency of
the preheating process~\cite{Chacko:2002wr}. The possible source of
the nontrivial orbit of the inflaton in our case is the $A$-term such
as
\beq
V_A  = a_u y_u m_{\tilde{q}} \Phi^3 + {\rm h.c.},
\eeq
where $a_u$ is a constant of $O(1)$.  The effect of this $A$-term is,
however, so small that the inflaton $\Phi$ exhibits almost
straight-line motion on the complex plane. Therefore, when the
inflaton comes closest to the origin, its amplitude is much smaller
than the critical value, below which the adiabaticity condition is
violated. Hence the preheating should proceed in the same way as a
real scalar field. In fact we have confirmed numerically that the
instability band almost coincides with that in the case of a real
scalar field, even in the presence of the $A$-term. Also, since the
preheating occurs very efficiently and ceases within several
oscillations as shown below, cumulative disturbance of the homogeneous
motion caused by the $A$-term can be neglected.

We would like to focus on the four-point scalar interaction with the
stop among many interactions the inflaton feels. When the inflaton
$\phi$ first reaches $\phi=0$, the particle production occurs and the
stops are generated, typically with the momentum $k_{\rm res}\sim
y_t^{1/2}\lambda^{1/4}\phi_0$ and the occupation number $n_k\sim O(1)$
for $k\sim k_{\rm res}$, where $\phi_0$ is the amplitude of the
oscillation. Since the stop obtains an effective mass $m_{\tilde{t}}
\sim y_t |\phi|$ through the four-point interaction, generated stops
are fattened as the amplitude of inflaton increases, and become as
heavy as $\sim y_t M_*$ around the endpoint of the oscillation of the
inflaton. The stop can then decay into two fermions, wino and
(left-handed) bottom quark, for example, because the time scale of the
decay is much shorter than that of the oscillation of the inflaton.
Similar process occurs when $\phi$ reaches $\phi=0$ again.  Each time
$\phi$ passes its origin, stops are generated, and they decay when
$\phi$ reaches around its maximal value.  This type of preheating is
known as ``instant preheating'' \cite{inst}. Applying the result of
Ref.~\cite{inst} to our case, the amount of the energy dissipated by
each decay of the stops is comparable to that originally stored in the
inflaton. Thus, the decay processes of the inflaton proceed very
efficiently and complete within several oscillations.

After that, the decay products are quickly thermalized through decays,
scatterings, and annihilations by gauge interactions, which proceed
very efficiently.  The reheating temperature $T_{RH}$ is expected to
be very high: $T_{RH} \sim 0.1 \lambda^{1/4} M_* \sim 10^{14} {\rm
GeV}$. Such high reheating temperature might lead to the
overproduction of dangerous relics like gravitinos~\cite{Khlopov:pf}.
In order to avoid this problem, we assume that one of the followings
is realized.  One solution is to assume large late-time entropy
production from, for example, thermal inflation~\cite{Lyth:1995ka}.
Another is to have a relatively heavy gravitino mass, $m_{3/2}
\simeq 10 - 100 {\rm TeV}$, so that the gravitinos can decay well
before the big bang nucleosynthesis (BBN) epoch, $T_{\rm BBN}\sim
1{\rm MeV}$. However, this does not remedy the situation if the
lightest supersymmetric particle (LSP) is the standard bino-like LSP,
since those produced from the decay of gravitinos would overclose the
universe. This difficulty can be evaded by the introduction of a
supersymmetric partner with a mass much lighter than $100{\rm GeV}$.
One possibility may be the axino, superpartner of the axion
\cite{Asaka:2000ew}.

So far we have not mentioned how the flat direction can take the value
of $O(M_*)$ or larger avoiding the Hubble-induced mass term that
prevents the flat direction from slow-rolling (so called
$\eta$-problem).  In the minimal supergravity model, the scalar
potential includes an exponential factor which essentially precludes
any scalar amplitudes larger than $M_*$, and scalar masses of the
order of $H$ are generated.  An idea to circumvent these obstacles to
construct a successful chaotic inflation model is to introduce the
Heisenberg symmetry~\cite{Binetruy:1987xj} under which the inflaton
$\phi$ and a chiral field $z$ transform as follows,
\beq
\delta z = \epsilon^* \phi,~~~
\delta \phi = \epsilon,
\eeq
where $\epsilon$ is a complex parameter.  We can construct an invariant
combination $y$ from $z$ and $\phi$ as
\beq
y \equiv z+z^*- \phi^* \phi.
\eeq
Imposing the Heisenberg symmetry in the K\"ahler potential, it is
written only with $y$, {\it i.e.}, $K=f(y)$.  It is easy to see that
$y$ and $\phi$ should be regarded as independent variables since the
kinetic terms are diagonalized for these
variables~\cite{Gaillard:1995az}.  It was shown that this symmetry
protects the flatness of the inflaton potential from both the
exponential growth and Hubble-induced mass term.  It should be
emphasized that the introduction of a new degree of freedom $y$ is the
price for keeping the potential flat.  Central to this issue is the
problem how to stabilize $y$. The dynamics of $y$ with a specific form
of $f(y) = \frac{3}{8}\ln y+y^2$ was discussed in
Ref.~\cite{Murayama:xu}, and it was found that the value of $y$ is
fixed during the inflation. In addition, another mechanism of fixing
$y$ using the radiative corrections was proposed in
Ref.~\cite{Gaillard:1995az} in the case of the no-scale supergravity
model~\cite{Cremmer:1983bf} where the K\"ahler potential takes a
special form as $K=-3\ln y$. In both cases, the potential of the
inflaton is same as that in the global supersymmetry case as long as
$y$ is fixed.  In this letter, we just assume $y$ is somehow fixed and
remains constant for successful chaotic inflation.\footnote
{Notice that, even in this framework, soft supersymmetry breaking
terms required for our mechanism can be generated by introducing a new
supersymmetry breaking field $x$ (with a slight modification of the
K\"ahler potential).  For example, let us consider the K\"ahler
potential of the form $K=f(y)+a_{ij}|x|^2(\phi_i^*\phi_j+{\rm h.c.})$
with $a_{ij}$ being constants, $\phi_i$ MSSM chiral multiplets, and
$y=z+z^*+\phi_i^*\phi_i+|x|^2$.  This K\"ahler potential does not have
the Heisenberg symmetry if $a_{ij}\ne 0$.  Previous arguments,
however, still hold since $|x|\ll M_*$ is realized during the
inflation because of the Hubble-induced mass of $x$; with $|x|\ll
M_*$, inflaton potential does not change.  In addition, if the
$F$-component of $x$ is non-vanishing (and large enough) in the
vacuum, soft supersymmetry breaking scalar masses squared (including
the flavor-violating ones) are generated.  Although the $F$-component of 
$x$ contributes to the cosmological
constant, it is possible to cancel it out with a fine-tuning of the
Kahler potential.  If we consider the model given in
Ref.~\cite{Murayama:xu}, for example, vanishing cosmological constant
can be realized by a rescaling of the Kahler potential.
Gaugino masses can arise
from $x$-dependent gauge kinetic functions (or from direct coupling of
$z$ to the gauge kinetic terms as suggested in
Ref.~\cite{Murayama:xu}).  $A$-parameters are also generated by
introducing $x$-dependent higher-dimensional terms in the
superpotential (or by the renormalization-group effects).}

Finally, we comment on the case with the $R$-parity violation.  In the
MSSM, there are $D$-flat directions which are not lifted by the Yukawa
interactions given in Eq.\ (\ref{W}).  Such flat directions are
parameterized by the monomial of the superfields with odd $R$-parity,
and can be lifted if the $R$-parity is broken.  Choosing relevant
$D$-flat direction, $R$-parity breaking Yukawa coupling can be as
large as $\sim O(10^{-6}-10^{-7})$ without conflicting experimental
bounds if the baryogenesis takes place after the sphaleron interaction
becomes ineffective \cite{Campbell:1990fa}.  Thus, if we adopt
$R$-parity violation of this size, such an $R$-parity violating
$D$-flat direction can be another candidate of the inflaton.  The
decay of $R$-parity violating $D$-flat directions occurs in the same
way as the previous case via the ``instant preheating.''  Since they
do not necessarily have interactions with stops, the four-point scalar
interaction with the largest coupling constant is expected to come
from the $D$-term. In particular, there are four-point scalar
interactions $\sim g^2 \chi^2 \phi^2$, where $\chi$ represents the
field orthogonal to the flat direction.

In summary, we have investigated whether the inflation can be embedded
in the MSSM sector, and found that the $D$-flat direction consisting
of the first generation left- and right-handed up squarks and the
up-type Higgs boson may be the inflaton if the up-quark mass
predominantly comes from the one-loop threshold correction to the
up-Yukawa coupling constant. The dynamics of the inflaton is almost
the same as the $\lambda \phi^{4}$ model, which has attracted much
attention recently. Since the inflaton in our model consists of the
MSSM particles, it is not only minimal but also testable at
high-energy collider experiments.  In particular, if the $QUH_u$ flat
direction plays the role of the inflaton, mixings of the up- and
top-squarks should be large.  This is an interesting check point of
our model and can be tested by collider experiments as well as precise
measurements of flavor-changing processes.

{\sl Acknowledgments:} 
We would like to thank T. Yanagida for noticing us the earlier work on the
similar subject.
F.T. is grateful to M. Kawasaki for useful
discussion.  The work of T.M. is supported by the Grant-in-Aid for
Scientific Research from the Ministry of Education, Science, Sports,
and Culture of Japan, No.\ 15540247.  F.T. thanks the Japan Society
for the Promotion of Science for financial support.

\end{document}